\documentclass[doublecol,figures]{epl2}
\pdfoutput=0
\newcommand{\Hamil}{\mathcal{H}}
\newcommand{\dg}{\dagger}
\newcommand{\pd}{{\phantom{\dagger}}}

\newcommand{\avg}[1]{\ensuremath{\langle #1 \rangle}}

\usepackage{amsmath}
\usepackage{braket} 
\usepackage{comment} 

\title{Going beyond the linear approximation in describing
  electron-phonon coupling: relevance for the Holstein model}

\shorttitle{Going beyond the linear electron-phonon coupling
approximation}

\author{C. P. J. Adolphs\inst{1} \and M. Berciu\inst{1,2}}

\institute{
\inst{1}Department of Physics and Astronomy, University of
  British Columbia, Vancouver BC V6T 1Z1, Canada

\inst{2}Quantum Matter Institute, University of
  British Columbia, Vancouver BC V6T 1Z4, Canada
}


\abstract{
Using the momentum average approximation we study the
  importance of adding higher-than-linear terms in the electron-phonon
  coupling on the properties of single polarons described by a
  generalized Holstein model.  For medium and strong linear coupling,
  even small quadratic electron-phonon coupling terms are found to
  lead to very significant quantitative changes in the properties of
  the polaron, which cannot be captured by a linear Holstein
  Hamiltonian with renormalized parameters. We argue that the
  bi-polaron phase diagram is equally sensitive to
  addition of quadratic coupling terms if the linear coupling is
  large. These results suggest that the linear approximation
is likely to be inappropriate to model
systems with strong electron-phonon coupling, at least for low carrier
concentrations.}

\begin{document}

\maketitle

\section{Introduction}

Coupling of carriers to phonons and the properties of the
resulting quasiparticles, the polarons, are important for many
materials, {\em e.g.} organic semiconductors \cite{org}, cuprates
\cite{cupr}, manganites \cite{mang}, two-gap superconductors like
MgB$_2$ \cite{MgB2}, etc.  In some cases the effective electron-phonon
(el-ph) coupling $\lambda$ is known quite accurately. For others, like
the cuprates, estimates range from very small ($\lambda \sim 0.3$) to
very large ($\lambda \sim 10$) \cite{cup2}. One possible explanation
for this is that, especially for stronger couplings where simple
perturbational expressions are no longer valid, properly fitting the
experimental data to theoretical models can be quite involved
\cite{ARPES}.

Here we consider another possible explanation, namely that
at strong el-ph coupling, simple theoretical models
may not be valid anymore. All widely-used models \cite{Holstein,Fro}
assume at the outset that the displacements $x_i$ of the atoms out of
equilibrium are small enough to justify expanding the electron-lattice
interactions to linear order in $x_i$. These linear models generically
predict the formation of small polarons or bipolarons at strong
coupling, with the carrier(s) surrounded by a robust phonon cloud. As
a result, lattice distortions $\langle x_i\rangle$ are considerable
near the carrier(s). Hence, the linear models are based on
  assumptions which are in direct opposition to their
  predictions.

In this Letter we investigate this issue in the {\em single polaron limit},
relevant for the study of weakly doped materials like very underdoped
cuprates \cite{cupr2} and organic semiconductors \cite{org}, and for
cold atoms/molecules simulators \cite{ca}.  We
study the ground-state (GS) of a single polaron in a generalized
Holstein model including el-ph coupling up to quartic order in $x_i$
to test the importance of the higher order terms.  We find that for
strong linear coupling even very small quadratic terms {\em
drastically} change the properties of the polaron. Moreover, we show
that these effects go beyond a mere renormalization of the parameters
of the linear Holstein model. As a result, attempts to find effective
parameters appropriate for a linear model by using its predictions to
fit the properties of real systems are doomed to failure, as
different values will be obtained from fitting different
properties. This offers another possible explanation for the wide
range of estimates of the el-ph coupling in some
materials. More importantly, it means that we must seriously
reconsider how to characterize such interactions when they are
strong. Furthermore, this calls for similar investigations of the
validity of these linear
models at finite carrier concentrations, since it is reasonable to
expect that they also fail in the strong coupling limit.

To the best of our knowledge, we present here the first systematic,
non-perturbative study of the importance of higher-order
el-ph coupling terms on single polaron properties.
We note that in  previous work going beyond linear
models, purely quadratic (no
linear term) but weak el-ph coupling was discussed for organic metals
using perturbation theory \cite{ora}, while linear and quadratic el-ph
coupling  was studied
in the context of high-T$_{\rm
  C}$ superconductivity in Ref.~\cite{crespi}. A semi-classical study
of some non-linear coupling potentials was carried out in Ref.~\cite{kenkre}.

\section{Formalism}
\begin{figure}[t]
\centering  \includegraphics[width=0.8\columnwidth]{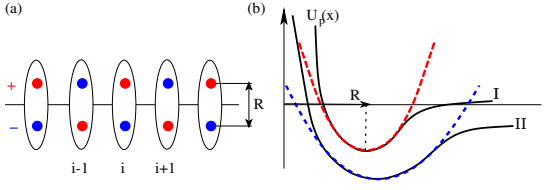}
  \caption{(a) Sketch of a 1D chain of polar molecules;
    (b) The potential of the pair with (II) or without (I) an extra charge carrier
    (full lines) is approximated by a polynomial (thick
    dashed lines)}
  \label{fig1sup}
\end{figure}

We use the momentum average (MA) approximation to carry out this
study.  MA was  shown to be very accurate in describing GS polaron properties for the
linear Holstein model, where it satisfies exactly multiple sum rules and becomes
asymptotically exact in the limit of strong coupling \cite{MA0}.  It is
straightforward to verify that all these considerations remain valid for the
generalized Holstein model:
$\Hamil = {\cal H}_{\rm el} + {\cal H}_{\rm ph} +
    {\cal H}_{\rm el-ph}$,
defined as follows. The Holstein Hamiltonian models
a charge carrier in a molecular crystal like the 1D example sketched
in Fig.~\ref{fig1sup}(a).
A charge carrier
introduced in such a crystal hops between
``molecules'', as described by  ${\cal H}_{\rm el}=\sum_{\vect{k}} \epsilon_{\vect{k}} c^\dg_{\vect{k}}
c^\pd_{\vect{k}}$ with $\epsilon_{\vect{k}} = - 2t
\sum_{\alpha=1}^{d} \cos(k_\alpha)$  for
nearest-neighbor hopping on a $d$-dimensional
simple cubic lattice.
Fig.~\ref{fig1sup}(b) illustrates how the lattice part is handled. In the
absence of a carrier, the potential has some form (curve I) which is
approximated as a parabola and leads to
$ {\cal
  H}_{\rm ph}= \Omega \sum_i b_i^\dg
b_i^\pd$.
This describes harmonic oscillations  of each
``molecule'' about its equilibrium
distance $R$. If a carrier is present, the potential has some other form (curve
II). The difference between I and II leads to ${\cal H}_{\rm el-ph}$. Its details are material specific;
here we propose two models and choose a generic form based on them.

The first model assumes that the  carrier
occupies an orbital of the ion with opposite charge. The
attraction between them is then some constant, whereas the
Coulomb  repulsion between the carrier and the ion of like charge
is
$
U(x_i) = \frac{U_0n_i}{1- x_i/R}= U_0 n_i \sum_{n=0}^{\infty}
\left(\frac{x_i}{R}\right)^n
$
where $n_i=1 (n_i=0)$ if the carrier is (is not) present and $U_0 >0$ is
the characteristic energy. Using $x_i = \sqrt{\frac{\hbar}{2\mu
  \Omega}} (b_i + b_i^\dagger)$ where $\mu$ is the reduced mass of the
molecule, and truncating the series at
$n=4$ leads to:
\begin{equation}
  \label{eq:el-ph}
{\cal H}_{\rm el-ph} = \sum_{n=1}^{4}  {\cal H}_{\rm el-ph}^{(n)}=
\sum_{n=1}^{4} g_n \sum_{i}^{} c_i^\dagger c_i (b_i + b_i^\dagger)^n,
\end{equation}
where $g_n = g_1 \zeta^{n-1}$ with  $\zeta = A/R $ and
$A=\sqrt{\hbar/(2\mu\Omega)}$ the zero-point amplitude of the
 harmonic oscillator.

The second  model assumes that the carrier is an electron (hole)
that occupies an anti-bonding (bonding) orbital of the molecule;  all
bonding orbitals are initially full  since the parent crystal is an
insulator.  In both cases the energy  increases by
an overlap integral which decreases exponentially with the distance:
 $U(x_i) \sim  n_i  e^{-\frac{R-x_i}{a_B}}$
where $a_B$ is the Bohr radius.
A Taylor expansion to fourth order in $x_i$ leads again to
Eq.~(\ref{eq:el-ph}) but now $g_n/g_1= 2^{n-1}
\zeta^{n-1}/n!$ for $\zeta=g_2/g_1= A/ (2a_B)$, where again
$A=\sqrt{\hbar/(2\mu\Omega)}$.

We define as the \emph{linear model} the case where only $g_1\ne 0$
(i.e., the usual Holstein model); as the \emph{quadratic model} the case where
only $g_1\ne 0, g_2\ne 0$; and as the \emph{quartic model} the case
where all  $g_n\ne 0$.  The  case with only $g_4=0$ is not
considered because it is unstable.

The linear Holstein model is characterized by two dimensionless
parameters: the effective coupling strength $\lambda = g_1^2 /
(2dt\Omega)$, where $d$ is the dimension of the lattice, and the
adiabaticity ratio $\Omega/(4dt)$.  As long as the latter is not very
small, the former controls the phenomenology, with the crossover to
small polaron physics occurring for $\lambda\sim 1$ \cite{Hrev}.  For
ease of comparison, we continue to use these parameters when
characterizing the higher order models.  For the quadratic model, the
new energy scale $g_2$ results in a third dimensionless parameter
$\zeta = g_2/g_1$.  For the quartic model there are two more
  parameters $g_n/g_1$, $n=3,4$.  Both scale like $\zeta^{n-1}$ but
  with different prefactors. We use $g_n/g_1=\zeta^{n-1}$ like in the
  first model since for the second model the prefactors are less than
  1, making these  terms smaller and thus less important.

For specificity, from now we assume $\zeta >0$
($\zeta <0$  is briefly discussed at the end). As we show below, in
this case we find that while quadratic terms are important when
the linear coupling is large, addition of the $n=3,4$
terms only leads to small quantitative changes and can be ignored. This justifies a
posteriori why
we do not include anharmonic corrections in ${\cal H}_{\rm ph}$ and/or
higher order terms with $n > 4$ in the electron-phonon coupling.

We now describe in detail the MA solution for the quadratic model.  The
calculations for the quartic model are analogous  but much more tedious.

We want to find the single particle Green's function $G(\vect{k},\omega) =
\braket{0| c^\pd_{\vect{k}} \hat G(\omega) c^\dg_{\vect{k}} | 0 }$ where $\hat
G(\omega)=[\omega - \mathcal{H} + i\eta]^{-1}$ is the resolvent for this
Hamiltonian, with $\eta \rightarrow 0$ a small positive number and $\ket{0}$ the
vacuum state. From this we can
extract all the polaron's GS properties \cite{MA0}.  We rewrite the
quadratic Hamiltonian as
$\mathcal{H} = \mathcal{H}_0 + \mathcal{H}_1$, where $\mathcal{H}_0=
{\cal H}_{\rm el} + {\cal H}_{\rm ph}
  +g_2 \sum_i c_i^\dg c_i^\pd \left(2b_i^\dg b_i^\pd+1\right)
$
while $\mathcal{H}_1= \sum_i c_i^\dg c_i^\pd \left[ g_1 (b_i^\dg + b_i^\pd)
    + g_2 (b_i^{\dg 2} + b_i^2) \right].$
The equation of motion (EOM) for the propagator is obtained recursively from
Dyson's identity, $\hat G(\omega) = \hat G_0(\omega) + \hat G(\omega) \Hamil_1
\hat G_0(\omega)$ where $\hat G_0(\omega) = [\omega - \Hamil_0 + i\eta]^{-1}$ is
the resolvent for ${\cal H}_0$.  Using it in $G(\vect{k},\omega)$ yields the EOM
\begin{equation}
  \label{eq:Geom}
G(\vect{k},\omega) = G_0(\vect{k},\omega) \left[ 1+\sum_{n=1}^{2}\sum_i \frac{e^{i
\vect{k}\cdot
      \vect{r}_i}}{\sqrt{N}} g_n F_n(\vect{k},\omega;i) \right]
\end{equation}
where  $
  F_n(\vect{k},\omega;i) = \braket{0 | c^\pd_{\vect{k}} \hat G(\omega) c_i^\dg
    (b_i^\dg)^n |0}.$

Applying Dyson's identity to generate EOM for the
$F_n$ propagators results in an infinite system of
coupled equations which involves many other generalized
propagators.  MA \cite{MA0,MAh}
circumvents this complication by making the
approximation $G_0(i-j,\omega;n) \approx \delta_{ij}
\bar g_0(\omega;n)$ for any $n\ge 1$, where
$G_0(i-j,\omega;n)=\frac{1}{n!}\braket{0 |
  c_i^\pd b_i^n \hat G_0(\omega) (b_i^\dg)^n
    c_i^\dg | 0} $.  This is justified
because the polaron GS energy lies below the free particle spectrum,
and for such energies
the free-particle propagator  decreases exponentially with
$|i-j|$.  Thus, MA keeps the largest contribution and ignores the
exponentially smaller ones.  This becomes exact in the
strong-coupling limit $t\rightarrow 0$.  The
propagator
$ \bar g_0(\omega;n)=\left[1/
     \tilde g_0(\omega - n\Omega - g_2)
      - 2g_2 n\right]^{-1}$
is that of a carrier scattered by an on-site potential $2g_2
n$, where  $\tilde g_0(\omega) = \frac{1}{N}
\sum_{\vect{k}}^{}  1/(\omega -
  \epsilon_{\vect{k}} +i \eta)$.

MA allows us to obtain a simplified hierarchy of EOM involving only
the generalized Green's functions $F_n$. For any
$n\ge 1$, they read:
  \begin{align}
    \label{eq:Feom}
  &  F_n(\vect{k},\omega;i) = \bar g_0(\omega;n) \cdot \big[
    n(n-1) g_2 F_{n-2}(\vect{k},\omega;i) \notag\\
    &+ n g_1 F_{n-1}(\vect{k},\omega;i) + g_1 F_{n+1}(\vect{k},\omega;i)
    + g_2 F_{n+2}(\vect{k},\omega;i) \big]. \nonumber
  \end{align}
Since the arguments of all $F_n$ propagators are the same, we
suppress them in the following for simplicity.

Following the technique introduced in Ref.~\cite{moller}, we reduce
this to
a simple recursive relation for the vector $\vect{W_n} =
(F_{2n-1}, F_{2n})$.  The EOM for $\vect{W_n}$ are
$ \tens{\gamma_n} \vect{W_n} = \tens{\alpha_n} \vect{W_{n-1}} + \tens{\beta_n} \vect{W_{n+1}},$
where the $\tens{\alpha_n}$, $\tens{\beta_n}$ and $\tens{\gamma_n}$ are $2 \times 2$
matrices whose coefficients are read off of the EOM, namely $
\tens{\alpha_n}|_{11} =(2n-1)(2n-2) g_2 \bar g_0(\omega; 2n-1)$,
$\tens{\alpha_n}|_{12} = (2n-1) g_1 \bar g_0(\omega; 2n-1)$,
$\tens{\alpha_n}|_{21} = 0$ and
$\tens{\alpha_n}|_{22} = 2n(2n-1) g_2 \bar g_0(\omega; 2n)$, while
\begin{equation}
\tens{\beta_n} =
  \begin{pmatrix}
    g_2 \bar g_0(\omega; 2n-1) & 0 \\
    g_1 \bar g_0(\omega; 2n) & g_2 \bar g_0(\omega; 2n)
  \end{pmatrix},
\end{equation}
\begin{equation}
  \tens{\gamma_n} = \begin{pmatrix}
    1 & -g_1 \bar g_0(\omega; 2n-1) \\ -2n g_1 \bar g_0(\omega; 2n) & 1
  \end{pmatrix}.
\end{equation}

This simple recursive relation for $\vect{W_n}$ has the solution
$\vect{W_n} = \tens{A_n} \vect{W_{n-1}}$ for any $n\ge 1$, where $\tens{A_n}$ are $2 \times 2$
matrices obtained from the infinite continued fraction
\begin{equation}
  \label{eq:7}
  \tens{A_n} = \left[ \tens{\gamma_n} - \tens{\beta_n} \tens{A_{n+1}} \right]^{-1}
  \tens{\alpha_n}.
\end{equation}
In practice, we start with $A_N=0$ for a sufficiently large cutoff
$N$, chosen so that the results are insensitive to further increases in
it ($N\sim 100$ is usually sufficient).

We find  $\tens{A_1}= \begin{pmatrix}0 & a_{12} \\ 0 & a_{22}
\end{pmatrix}$, where $a_{12}$ and $a_{22}$ are obtained after using
Eq.~(\ref{eq:7}) $N-1$ times.  As a result,  $F_1 = a_{12} F_0$, $F_2 = a_{22}
F_0$, where
$ G(\vect{k},\omega) = \sum_i e^{i\vect{k}\cdot\vect{r}_i}/\sqrt{N}
F_0(\vect{k},\omega;i)$.  Using these in Eq.~(\ref{eq:Geom}) leads to
a solution of
the expected form $G(\vect{k},\omega) = \left[ \omega -
  \epsilon_{\vect{k}} - \Sigma(\omega) +
  i\eta\right]^{-1}$, with the MA self-energy for the quadratic model:
$\Sigma(\omega) = g_1 a_{12}(\omega) + g_2 a_{22}(\omega)$.

The reason why the self-energy  is local at this level of
MA is the simplicity of this Hamiltonian, whose vertices are momentum
independent; this issue is discussed at length for the linear Holstein model in
Ref.~\cite{MAh}.

The quartic model is solved analogously.  The main difference is that here the
EOM for $F_n$ involves 9 consecutive terms, from $F_{n-4}$ to $F_{n+4}$.  These
can also be rewritten as simple recurrence relations $\tens{\gamma_n} \vect{W_n}
= \tens{\alpha_n} \vect{W_{n-1}} + \tens{\beta_n} \vect{W_{n+1}},$ but now
$\tens{\alpha_n}$, $\tens{\beta_n}$ and $\tens{\gamma_n}$ are $4\times 4$ matrices.
Their expressions are too long to be listed here.

\section{Results and Discussion}

\begin{figure}
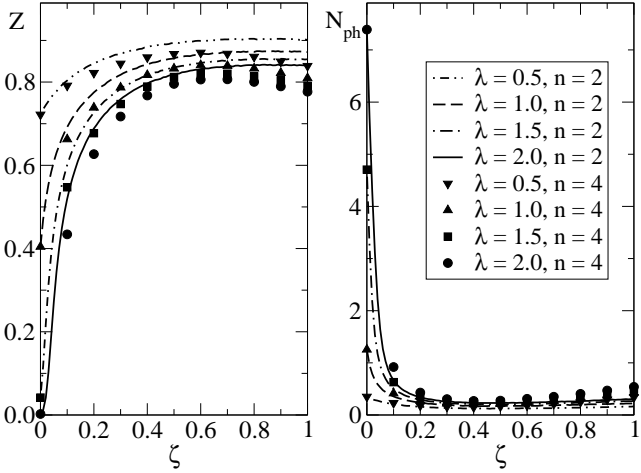

  \onefigure[width=0.95\columnwidth]{fig2.eps}
  \caption{GS quasiparticle weight (left panel) and GS
    average phonon number (right panel) vs.~$\zeta$, in the quadratic
    ($n=2$, lines) and quartic ($n=4$, symbols) models, for various
    values of $\lambda$ and $\Omega = 0.5t$, in one dimension.
  \label{fig:zeta_lines}}
\end{figure}

To gauge the relevance of the higher-order el-ph coupling
  terms we plot in Fig.~\ref{fig:zeta_lines} the evolution with
  $\zeta$ of  a polaron property that can be directly
measured, namely the quasiparticle weight
$Z = m/m^*$ where $m, m^*$ are the carrier and the polaron
mass, respectively. We also show the average phonon number
$N_{ph}$. The results are for a one-dimensional chain. Results in
higher dimensions are
qualitatively similar to these 1D results for small $\lambda$, and become
quantitatively similar to them in the interesting regime of large
$\lambda$ where all of them converge towards those of the atomic limit $t=0$.

First, we note that the $\zeta=0$ intercepts trace the predictions of the linear
model: with increased coupling $\lambda$, $Z$ decreases while $N_{ph}$ increases
as the polaron acquires a robust phonon cloud \cite{MA0,Hrev}.  From these
intercepts, we estimate that the linear model predicts the crossover to the
small polaron regime to occur around $\lambda\sim 1.5$ for this adiabaticity
ratio and dimension.

The quadratic model, whose predictions are indicated by lines, shows
a very strong dependence of $\zeta$ for strong linear
coupling $\lambda \ge 1.5$: here both $Z$ and $N_{ph}$ vary by about
an order of magnitude as $\zeta$ increases from 0 to $0.1$. For higher
$\zeta$, $Z$ and $N_{ph}$ have a slight turnaround towards
smaller/larger values, for reasons explained below, but are still
consistent with a large polaron.  These results indicate
that the quadratic term can completely change the behavior of the
polaron in the limit of medium and large $\lambda$.  For example, in
the quadratic model at $\lambda=1.5$ and $\zeta\sim 0.1$ the polaron
is light and with a small phonon cloud, in total disagreement with the
linear model prediction of a heavy small polaron at this $\lambda$.

Of course, this raises the question of how large $\zeta$ is.
 The answer is material specific, but as an extreme case,
 let H$_2$ be the unit of the molecular crystal. This case is
 described by model two, so $\zeta \sim 2A/a_B$, where $a_B
 \approx 0.5\AA$ while $A\approx 0.1\AA$ if we use $\Omega \approx
 0.5eV$ appropriate for a H$_2$ molecule \cite{Herz}. This leads to a
 very large $\zeta \sim 0.4$. Other atoms are heavier but phonon
 frequencies are usually much smaller than $0.5eV$, so it is not clear
 whether $A\sim 1/\sqrt{\mu \Omega}$ changes much. The Bohr radius (or
 distance $R$ between atoms, for model 1) is usually larger than
 $0.5\AA$ but not by a lot, maybe up to a factor 5 for $R$; thus we
 expect smaller $\zeta$ in real materials but the change is likely
 not by orders of magnitude. Fig.~\ref{fig:zeta_lines} shows that
 values as small as $\zeta\sim 0.05$ already lead to significant
 quantitative changes in $m^*$.

Inclusion of
cubic and quartic terms (the symbols show the results of the quartic model)
further changes $Z$ and $N_{ph}$, but these changes are much smaller for all
$\zeta$, of up to $\sim 10\%$ when compared to the quadratic model values, as
opposed to order of magnitude changes between the quadratic and the linear
models. Thus, these terms are much less relevant
  and can be ignored without losing much accuracy. As discussed,
  their small effect explains why we do not consider terms
  with even higher order $n$, nor $n=4$ anharmonic terms in the phonon
Hamiltonian.

To understand the effects of the quadratic term at large $\lambda$, we study it
in the atomic limit $t = 0$ ($\lambda=\infty$) where the carrier remains at one
site and interacts only with the phonons of that site.
Focusing on this site,
its quadratic Hamiltonian ${\cal H}^{(2)}_{\rm at}= \Omega b^\dg b +
\sum_{n=1}^{2} g_n (b^\dg + b)^n$ is well-studied in the field of quantum
optics, where it describes so-called squeezed coherent states
\cite{quant_opt}. The extra charge changes the origin and spring
  constant of the original harmonic oscillator which means that the Hamiltonian
  is easily diagonalized by
changing to new
bosonic operators $\gamma^\dg = u b^\dg + v b + w$, where $u$, $v$ and $w$ are
such that $\mathcal{H}^{(2)}_{\rm at} = \Omega_{\rm at}
\gamma^\dg \gamma + E^{(\text{at})}_{GS}$.
We find
 $ \Omega_{\rm at} = \sqrt{\Omega(\Omega+4g_2)}$,
  $u = \sqrt{ \left(\Omega +
        2g_2+\Omega_{\text{at}}\right)/(2\Omega_{\text{at}})}$,
  $w = g_1 \sqrt{\Omega/\Omega_{\text{at}}^3}$ and
  $v= \mathrm{sgn}(g_2) \sqrt{\left(\Omega +
      2g_2-\Omega_{\text{at}}\right)/(2\Omega_{\text{at}})}$.
From these, we obtain
 $ E_{\text{GS}}^{\text{at}} = -\frac{g_1^2\Omega}{\Omega_{\text{at}}^2} +
  \frac{1}{2} (\Omega_\text{at}  - \Omega)$,
$ N_{\text{ph}}^{(\text{at})} = \frac{1}{2} \left[ \frac{\Omega +
     2g_2}{\Omega_{\text{at}}} - 1\right] +
\frac{g_1^2}{(\Omega + 4g_2)^2}$ and $Z_{\rm at} = \frac{1}{u}
\exp\left[-w^2\left(1 - \frac{v}{u}\right)\right]$. The latter result
requires the expansion of
the squeezed coherent states in the number state basis \cite{squeezed}.

\begin{figure}
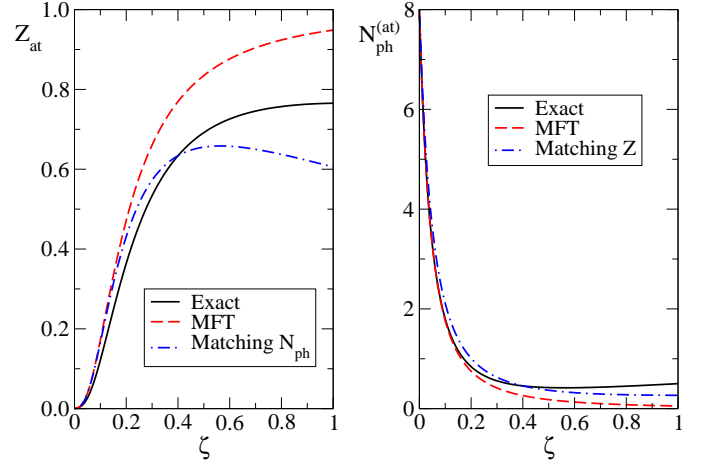

  \onefigure[width=\columnwidth]{fig3.eps}
  \caption{ (left) $Z_{\rm at}$, and (right)
	$N^{(\text{at})}_{\text{ph}}$ vs.~$\zeta$, for $g_1 = \sqrt{2}$ and
	$\Omega = 0.5$ (full lines). Dashed lines show the mean-field
	estimates, while the dot-dashed lines show the results of fitting
	$\tilde g / \tilde \Omega$ to exactly reproduce the other
	quantity. See text for more details.
  \label{fig:exact_vs_effective}}
\end{figure}

Figure~\ref{fig:exact_vs_effective} shows $Z_{\rm at}$ and
$N^{(\text{at})}_{\text{ph}}$ vs.~$\zeta$ (thick lines), which agree well
with the corresponding $\lambda=2$ results of Fig.~\ref{fig:zeta_lines}.
In particular, for $\zeta\rightarrow 0$ we find
$ \Omega_{\rm at} = \Omega + 2g_1\zeta+{\cal O}(\zeta^2)$, $
N_{\text{ph}}^{(\text{at})} = \frac{g_1^2}{\Omega^2}\left[1-{\frac{8g_1}{\Omega}\zeta+
\mathcal{O}}(\zeta^2)\right]$,
explaining their linear increase/decreases for small $\zeta$.

The slight turnaround of the $Z$ and $N_{\text{ph}}$ curves at
larger values of $\zeta$ is also observed in the atomic
limit of the quadratic model.  The reason is that the first term in
$N_{\text{ph}}^{\text{(at)}}$ \emph{increases}  whereas the
second term \emph{decreases} with $\zeta$.  As discussed above, for
small $\zeta$ the second term dominates and the overall number of
phonons decreases.  For large $\zeta$, however, the second term
vanishes whereas the first term diverges as $\sqrt{g_2}=\sqrt{\zeta
  g_1}$.  Hence, as $\zeta$ increases $N_{\text{ph}}^{\text{at}}$ has a
minimum, and then starts to increase with $\zeta$. Basically, here
the  $g_2(b^{\dagger 2} + b^2)$ coupling dominates over the linear coupling
$g_1(b^{\dagger} + b)$ and changes the trend.

This leads us to pose the question whether these exact results of the quadratic
atomic model can be fit well by an effective \emph{linear} model ${\cal
  H}^{(1)}_{\rm at} = \tilde{\Omega} b^\dg b + \tilde{g}(b^\dg + b)$, for some
appropriate choice of the effective parameters $\tilde{\Omega},\tilde{g}$.  One
way to achieve this is with a mean-field ansatz $b^{\dg 2} \approx 2 \avg{ b^\dg
} b^\dg -\avg{ b^\dg }^2$, with $\avg{b^\dg}$ the GS expectation value of
$b^\dg$.  The self-consistency condition $\avg{ b^\dg } = -(g_1 + 2g_2
\avg{b^\dg})/(\Omega + 2g_2)$ leads to the mean-field estimates
$\tilde{\Omega}_{\rm MF} = \Omega + 2 g_2$, $\tilde{g}_{\rm MF} = g_1 -
2g_1g_2/(\Omega + 4 g_2)$.  Thus, for small $\zeta=g_2/g_1$,
$\tilde{\Omega}_{\rm MF}$ increases whereas $\tilde{g}_{\rm MF}$ decreases with
increasing $\zeta$ so the effective coupling $\tilde{\lambda} =
\tilde{g}^2/(2dt\tilde{\Omega})$ decreases with $\zeta$.  This is consistent
with the observed move away from the small polaron limit with increasing
$\zeta$.  Quantitatively, however, these mean-field results (dashed lines in
Fig.~\ref{fig:exact_vs_effective}) are not very accurate for small $\zeta$, and
fail to capture even qualitatively the correct behavior when $\zeta\gg 1$, since
here $N^{(\text{at})}_{\text{ph}} \rightarrow \infty$ while
$N_{\text{ph}}^{(MF)}=\tilde{g}_{\rm MF}^2/\tilde{\Omega}_{\rm MF}^2 \rightarrow
0$.

In fact, there is no choice for effective linear parameters $\tilde{g}$ and
$\tilde{\Omega}$ that reproduces the results of the quadratic model.  This is
because in the linear model, both $\tilde{Z}$ and $\tilde{N}_{\text{ph}}$ are
functions of $\tilde{g}/\tilde{\Omega}$ only.  Fig.~\ref{fig:exact_vs_effective}
shows that if one chooses this ratio so that $N^{(\text{at})}_{\rm ph}=\tilde{N}_{\rm
  ph}$, then $\tilde{Z}$ (dot-dashed line in the left panel) disagrees with
$Z_{\rm at}$, and vice versa.  Even more significant is the fact that even if
one could find a way to choose $\tilde{g},\tilde{\Omega}$ so that the overall
agreement is satisfactory for all GS properties, the linear model's prediction
for higher energy features would still be {\em completely wrong}.  For example, it
would predict the polaron+one-phonon continuum to occur at
$E_{GS}+\tilde{\Omega}$ instead of the proper $E_{GS}+\Omega$ threshold.  Since
in the atomic limit the predictions of the quadratic model cannot be reproduced
with a renormalized linear model, we conclude that this must hold true at finite
hopping $t$ as well, at least for large $\lambda$ where the quadratic terms are
important.

So far we discussed moderate values of the adiabaticity ratio $\Omega/t=0.5$, as
well as the anti-adiabatic (atomic) limit.  MA predicts similar results in the
adiabatic limit $\Omega/t \rightarrow 0$ for large $\lambda$, where it remains
accurate, but is unsuitable to study small and moderate couplings
\cite{MAh}.  We expect that here the quadratic coupling is
essential even for small couplings $\lambda \rightarrow 0$, because
the term $2 g_2 \sum_{i}^{}b_i^\dg b_i$  insures that phonons  are
gapped even though $\Omega =0$.

So far we also only discussed the case $\zeta >0$.  The behavior of
models with $\zeta <0$ can be glimpsed at from the exact results  in
the atomic limit.  For
small negative $\zeta$, the results listed above show that the average
phonon number $ N_{\text{ph}}^{\text{(at)}}$ increases with $|\zeta|$ while the
{\em qp} weight $Z_{\rm at}$ decreases fast, i.e. the polaron moves more
strongly into the small polaron limit.  This is in agreement with the MA
predictions for the quadratic model (not shown). Here, however, we must limit
ourselves to values $|\zeta|<\Omega/(4g_1)$ so that $\Omega_{\rm at}$ remains a
real quantity (a similar threshold is found for the full quadratic model. Note
that the value of this threshold decreases with increasing $\lambda$).  For
values of $|\zeta|$ above this threshold the quadratic model becomes
unstable.  This, of course, is unphysical.  In reality, here one
is forced to include
higher order (anharmonic) terms in the phonon Hamiltonian ${\cal H}_{\rm ph}$
since they guarantee the stability of the lattice if the quadratic terms fail to
do so.  Such anharmonic terms may have little to no effect in
the absence of the carrier, but clearly become important in its presence, in
this limit. They can be treated with the same MA formalism we used here.  Their
effects, as well as a full analysis of all possible signs of the non-linearities
and the resulting polaron physics will be presented elsewhere.  For our current
purposes, it is obvious that in the case $\zeta <0$, higher order terms in el-ph
coupling also play a key role in determining the polaron properties unless
$\lambda$ is very small, and therefore cannot be ignored.

The results presented so far
clearly demonstrate the importance of non-linear el-ph coupling terms
if the linear coupling $\lambda$ is moderate or large, through their
significant effects on the properties of a single Holstein polaron.

A reasonable follow-up question is whether such dramatic effects are limited to
the single polaron limit or are expected to extend to finite carrier
concentrations.  While the limit of large carrier concentrations remains
to be investigated  in future work, here we present strong
evidence that quadratic terms are likely to be equally important at small but
finite carrier concentrations.

Of course, for finite carrier concentrations one needs to supplement the
Hamiltonian with a term describing carrier-carrier interactions.  The simplest
such term is an on-site Hubbard repulsion ${\cal H}_U = U \sum_{i}^{}
n_{i\uparrow} n_{i\downarrow}$, and gives rise to the Hubbard-Holstein
Hamiltonian.  The linear version of this Hamiltonian has been studied
extensively  by a variety of numerical
methods \cite{Hrev}.  In particular, for low carrier concentrations
and focusing on the small polaron/bipolaron limit, the  phase
diagram has been shown to consist of three regions: (i) for large $\lambda$ and
small $U$, the deformation energy favors the formation of on-site bipolarons,
also known as the $S_0$ bipolarons; (ii) increasing $U$ eventually makes having
two carriers at the same site too expensive, and the $S_0$ bipolarons evolve
into weakly-bound $S_1$ bipolarons, where the two carriers sit on neighboring
sites.  The binding is now provided by virtual hopping processes which allow
each carrier to interact with the cloud of its neighbor. However, at
smaller
$\lambda$ and larger $U$ this binding mechanism is insufficient to stabilize the
S1 bipolaron, and instead one finds (iii) a ground state consisting of unbound
polarons.

\begin{figure}
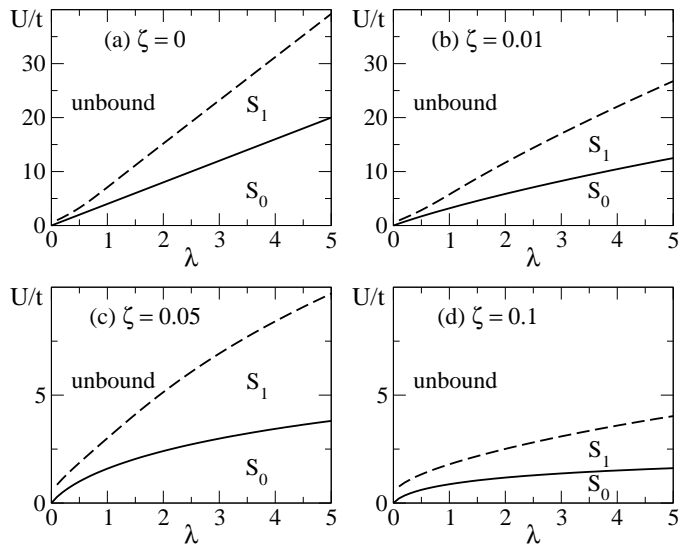

  \onefigure[width=\columnwidth]{fig4.eps}
  \caption{Estimate of the bipolaron phase diagram in 1D for
    $\Omega/t=0.5$ and for different
    values of $\zeta$, based on
second order perturbation theory in $t$. In all four panels,
the solid lines show the transition from $S_0$ (on-site) stable
bipolarons to $S_1$ (nearest-neighbor) stable bipolarons, while the
dashed lines show the unbinding transition above which bound polarons
are unstable. Note that panels (c) and (d) have a significantly
rescaled $y$-axis. See text for more details.
  \label{fig4}}
\end{figure}

This phase diagram
has been found numerically in 1D \cite{1D} and 2D \cite{2D} for the
linear Hubbard-Holstein model.  Some results in 3D have also become
available very recently \cite{3D}.  In 1D and 2D, the separation lines
between the various phases are found to be close to those estimated
using second order perturbation theory in the hopping $t$, starting
from the atomic limit \cite{1D,2D}.  This is expected since for large
linear coupling $\lambda$, the results always converge toward those
predicted by the atomic limit.

Since the quadratic Hamiltonian can be diagonalized exactly in the atomic limit,
we use second order perturbation theory to estimate the location of the
separation lines for various values of $\zeta>0$.  The results are shown in
Fig.~\ref{fig4}.  Panel (a) shows the rough phase diagram for $\zeta=0$, in
agreement with the asymptotic estimates shown in Refs.~\cite{1D,2D} (note that
the definition of the effective coupling used in those works differs by various
factors from our definition for $\lambda$).  Panels (b)-(d) show a very
significant change with increasing $\zeta$. Even the presence of an extremely small
quadratic term $\zeta=0.01$ moves the two lines to
considerably lower $U$ values, as shown in panel (b), while for $\zeta=0.05$ and
$0.1$, the bipolarons are stable only in a very narrow region with small values
of $U$ (note that the vertical axes are rescaled for panels (c) and (d)).

The dramatic change with increasing $\zeta$ in the location of these
asymptotic estimates for the various bipolaron transitions/crossovers
strongly suggests that non-linear el-ph coupling terms remain just as
important in the limit of small carrier concentrations as they have
been shown to be in the single polaron limit.  In particular, these
results suggest that the presence of non-linear el-ph coupling terms
leads to a significant suppression of the phonon-mediated interaction
between carriers, so that the addition of a small repulsion $U$
suffices to break the bipolarons into unbound polarons (whose
properties are also strongly affected by the non-linear terms, as
already shown).

The Holstein model is the simplest
example of a  $g(q)$ model, i.e. a model where the electron-phonon
interaction depends only on the momentum of the phonon.  Physically, such models
appear when the coupling to the lattice manifests itself through a modulation of
the on-site energy of the carrier.  The Fr\"ohlich model is another famous
example of $g(q)$ coupling.  Models of this type are found to have qualitatively
similar behavior, with small polarons forming when the effective coupling
increases.  These small polarons always have robust clouds, with significant
distortions of the lattice in their vicinity.  We therefore expect that
non-linear terms become important for all such models at sufficiently large
linear coupling.

To summarize, we have shown that non-linear terms in the el-ph coupling {\em
  must} be included in a Holstein model if the linear coupling is large enough
to predict small polaron formation, and that doing so may very significantly
change the results.  We also argued that these changes cannot be accounted for
by a linear Holstein model with renormalized parameters.  These results show
that we have to (re)consider carefully how we model interactions with phonons
(more generally, with any bosons) in materials where such interactions are
expected to be strong, at least for low carrier concentrations and for models
where this coupling modulates the on-site energy of the carriers.  Whether this
is also true in the metallic regime and/or for other types of models remains an
open question.

\acknowledgments
This work was supported by NSERC and QMI.

\end{document}